\documentclass[aip,apl,reprint,superscriptaddress,10pt]{revtex4-1}

\usepackage{graphicx}

\usepackage{amsmath}

\usepackage{color}

\usepackage{epstopdf}

\usepackage{multirow,bigstrut}

\usepackage{time,mathptmx,helvet}
\usepackage{lineno}

\usepackage[
pdfstartview=XYZ,
bookmarks=false,
colorlinks=true,
linkcolor=blue,
urlcolor=blue,
citecolor=blue,
pdftex
]{hyperref}
\usepackage{xcolor}
\usepackage{soul}
\makeatletter

\newcommand{\Rmnum}[1]{\expandafter\@slowromancap\romannumeral #1@}

\makeatother

\begin{document}
% \linenumbers
\title{\large\textcolor{blue}{Time-Resolved Interferometric Measurements of Plasma Density Evolution in Laser-Driven Capacitor-Coil Targets}\vspace{3pt}}

%\date{\today}

%---------------------------------------------------
%\linenumbers

\author{Yang~Zhang}
\affiliation{Department of Astrophysical Sciences, Princeton University, Princeton, NJ, 08544, USA} 
\affiliation{University Corporation for Atmospheric Research, Boulder, CO, 80301, USA }

\author{Ryo Omura}
\affiliation{Graduate School of Science, The University of Osaka, 1-1 Machikaneyama, Toyonaka, Osaka, 560-0043, Japan}
\affiliation{Institute of Laser Engineering, The University of Osaka, 2-6 Yamada-oka, Suita, Osaka, 565-0871, Japan}

\author{Rinya Akematsu}
\affiliation{Graduate School of Science, The University of Osaka, 1-1 Machikaneyama, Toyonaka, Osaka, 560-0043, Japan}
\affiliation{Institute of Laser Engineering, The University of Osaka, 2-6 Yamada-oka, Suita, Osaka, 565-0871, Japan}

\author{King Fai Farley Law}
\affiliation{Institute of Laser Engineering, The University of Osaka, 2-6 Yamada-oka, Suita, Osaka, 565-0871, Japan}

\author{Brandon~K.~Russell}
\affiliation{Department of Astrophysical Sciences, Princeton University, Princeton, NJ, 08544, USA} 

\author{Geoffrey~Pomraning}
\affiliation{Department of Astrophysical Sciences, Princeton University, Princeton, NJ, 08544, USA} 

\author{Kian Orr}
\affiliation{Department of Mechanical and Aerospace Engineering, Princeton University, Princeton, NJ, 08544, USA} 

\author{Kai Kimura}
\affiliation{Graduate School of Science, The University of Osaka, 1-1 Machikaneyama, Toyonaka, Osaka, 560-0043, Japan}
\affiliation{Institute of Laser Engineering, The University of Osaka, 2-6 Yamada-oka, Suita, Osaka, 565-0871, Japan}

\author{Muhammad Fauzan Syahbana}
\affiliation{Graduate School of Science, The University of Osaka, 1-1 Machikaneyama, Toyonaka, Osaka, 560-0043, Japan}
\affiliation{Institute of Laser Engineering, The University of Osaka, 2-6 Yamada-oka, Suita, Osaka, 565-0871, Japan}

\author{Yuga Karaki}
\affiliation{Graduate School of Science, The University of Osaka, 1-1 Machikaneyama, Toyonaka, Osaka, 560-0043, Japan}
\affiliation{Institute of Laser Engineering, The University of Osaka, 2-6 Yamada-oka, Suita, Osaka, 565-0871, Japan}

\author{Hiroki Matsubara}
\affiliation{Graduate School of Science, The University of Osaka, 1-1 Machikaneyama, Toyonaka, Osaka, 560-0043, Japan}
\affiliation{Institute of Laser Engineering, The University of Osaka, 2-6 Yamada-oka, Suita, Osaka, 565-0871, Japan}

\author{Ryuya Yamada}
\affiliation{Graduate School of Science, The University of Osaka, 1-1 Machikaneyama, Toyonaka, Osaka, 560-0043, Japan}
\affiliation{Institute of Laser Engineering, The University of Osaka, 2-6 Yamada-oka, Suita, Osaka, 565-0871, Japan}

\author{Jinyuan Dun}
\affiliation{Graduate School of Science, The University of Osaka, 1-1 Machikaneyama, Toyonaka, Osaka, 560-0043, Japan}
\affiliation{Institute of Laser Engineering, The University of Osaka, 2-6 Yamada-oka, Suita, Osaka, 565-0871, Japan}

\author{Ryunosuke Takizawa}
\affiliation{Institute of Laser Engineering, The University of Osaka, 2-6 Yamada-oka, Suita, Osaka, 565-0871, Japan}

\author{Yasunobu Arikawa}
\affiliation{Institute of Laser Engineering, The University of Osaka, 2-6 Yamada-oka, Suita, Osaka, 565-0871, Japan}

\author{Tatiana Pikuz}
\affiliation{Institute of Laser Engineering, The University of Osaka, 2-6 Yamada-oka, Suita, Osaka, 565-0871, Japan}

\author{Yuji Fukuda}
\affiliation{Kansai Photon Science Institute, National Institutes for Quantum and Radiological Science and Technology, 8-1-7 Umemidai, Kizugawa, Kyoto, 619-0215, Japan}

\author{Lan~Gao}
\affiliation{Princeton Plasma Physics Laboratory, Princeton University, Princeton, NJ, 08543, USA} 

\author{Hantao~Ji}
\affiliation{Department of Astrophysical Sciences, Princeton University, Princeton, NJ, 08544, USA} 
\affiliation{Princeton Plasma Physics Laboratory, Princeton University, Princeton, NJ, 08543, USA} 

\author{Shinsuke Fujioka}
\affiliation{Institute of Laser Engineering, The University of Osaka, 2-6 Yamada-oka, Suita, Osaka, 565-0871, Japan}

\begin{abstract}

\noindent 

Laser-driven capacitor-coil targets provide a compact platform for generating strong  magnetic fields and are widely used in magnetized high-energy-density plasma experiments. In addition to magnetic-field generation, these targets also produce plasma in the coil region, which can influence the subject physical processes, interact with secondary targets or external plasmas in their applications. However, direct, time-resolved measurements of the plasma density surrounding the coil remain limited. Here, we report interferometric measurements of the plasma density evolution in laser-driven capacitor-coil targets irradiated by the University of Osaka LFEX laser. Two-dimensional electron density maps reveal two distinct plasma sources loading the coil region: plasma generated in the coil itself and plasma produced by laser ablation of the target plates. These results provide quantitative information on plasma loading and evolution in capacitor-coil targets and are directly relevant to the design and modeling of magnetized high-energy-density plasma experiments.
\end{abstract}

%---------------------------------------------------

\pacs{}

%---------------------------------------------------

\maketitle

Laser-driven capacitor–coil targets are widely used to generate strong magnetic fields in high-energy-density (HED) physics and laboratory astrophysics \cite{Daido1986,Courtois2005,fujioka2013kilotesla,Santos_NJP_2015,Gao2016,Law2016,Goyon_PRE_2017,Peebles_PoP_2020,Chien21,Vlachos_PoP_2024,morita2023generation,Gao2025APL_CapacitorCoil,zhang2025accurate}. In a typical configuration, the target consists of two parallel plates connected by a conducting coil \cite{Gao2016}. When one plate is irradiated by a high-intensity laser pulse, superthermal electrons are produced in the laser–solid interaction and subsequently collected by the opposing plate, establishing a voltage difference between the two plates\cite{Fiksel_APL_2016} and then drive a current flowing through the coil\cite{PearlmanAPL,Forslund}.  For commonly used laser and target parameters, the peak current ranges from tens to hundreds kA, giving rise to quasi-static magnetic fields of tens to hundreds of tesla over millimeter-scale volumes. Owing to this capability, such targets have been adopted in magnetized HED experiments, including guiding and collimation of relativistic electron beams \cite{Bailly_NC_2018,Sakata_NC_2018}, control of hydrodynamic instabilities \cite{Matsuo_PRL_2021}, regulation of collisionless shock formation and hot-electron transport \cite{Woolsey_PoP_2001,Pisarczyk_PPCF_2022}, laboratory investigations of strongly driven magnetic reconnection \cite{chien2019,chien23,zhang23,yuan2023push,ji_PoP_2024}, and magnetized inertial confinement fusion for potential yield enhancement \cite{Perkins_PoP_2013}.

Despite extensive experimental and theoretical efforts devoted to measuring the magnetic fields generated by laser-driven capacitor-coil targets, the plasma density distribution in the vicinity of the coil has received comparatively little attention. While localized plasma parameters have been probed in a limited number of studies using Thomson scattering\cite{zhang23}, these measurements do not provide a global, time-resolved picture of the plasma environment surrounding the coil. Such information is essential, as the evolution of the self-generated plasma can influence both magnetic-field generation and its coupling to externally introduced plasmas in combined field-plasma experiments. In particular, plasma can be generated from the coil, and expansion from the laser interaction region may also load plasma in the coil area, interfering with application-region plasmas, potentially altering the underlying physical processes. Quantitative characterization of the plasma density evolution is therefore crucial for the reliable application and modeling of laser-driven capacitor-coil targets.

In this Letter, we present time-resolved interferometric measurements of the plasma density evolution in laser-driven capacitor–coil targets irradiated by the short-pulse LFEX laser at the University of Osaka at relativistic intensities. Two-dimensional electron density maps obtained at 1.0 and 3.1 ns after laser irradiation reveal two distinct plasma sources loading the coil region: a dense plasma localized near the coil and plasma supplied by expansion from the laser interaction region. To isolate the role of the coil, these measurements are directly compared with interferometry data from otherwise identical targets without a coil at the same time delays. In addition, proton radiography at 0.59~ns provides independent evidence of electric current generation. Together, these results offer new experimental insight into plasma loading and density evolution in laser-driven capacitor–coil targets and provide important constraints for their application in magnetized high-energy-density plasma experiments.

\begin{figure*}
\includegraphics[width=1\textwidth]{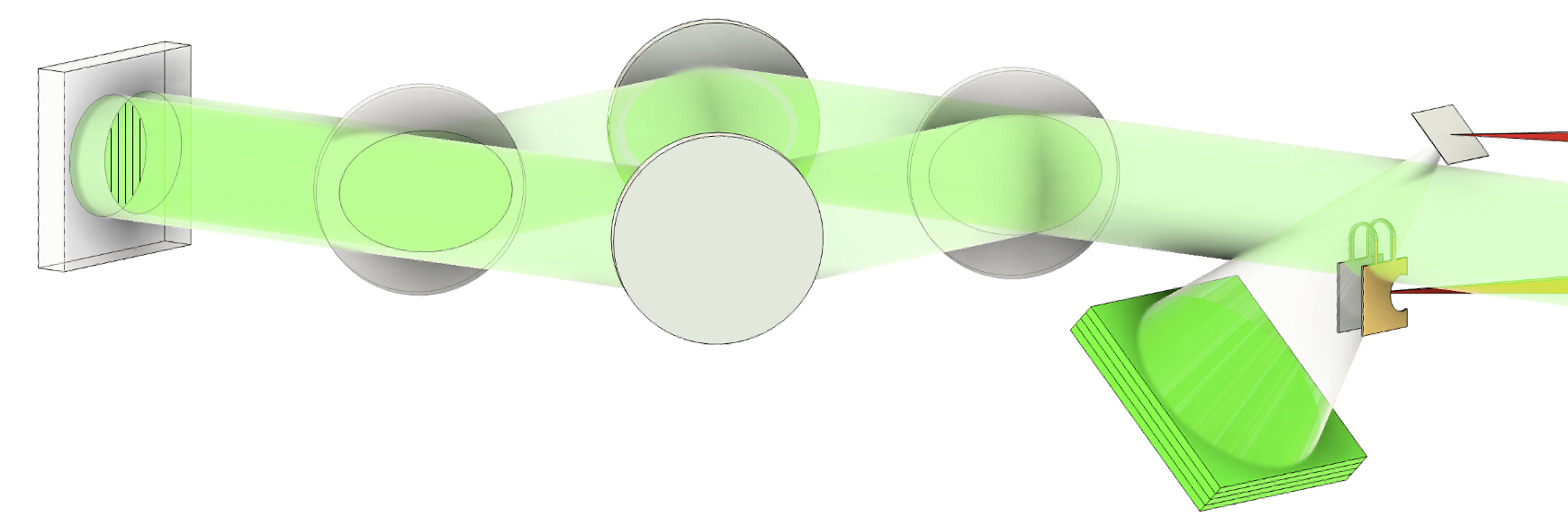}
\caption{\label{fig1} 
Experimental setup for interferometry and proton radiography. A probe laser beam with a diameter of 5 mm is used for interferometric measurements. The probe beam is much larger than the plasma region; therefore, two portions of the same beam are utilized, with one passing through the plasma and the other propagating outside the plasma as a reference. After transmission through the target, the beam is split and the plasma-affected portion is overlapped with the reference portion to produce interference fringes. Proton radiography is performed using a short-pulse laser irradiating a 10-$\rm{\mu}$m-thick and 1 mm $\times$ 1 mm aluminum foil target to generate protons via the TNSA mechanism. The proton beam propagates through the central region between the two coil peaks and is recorded on RCF stacks. The proton beam axis is oriented at 45° with respect to the vertical direction.
}
\end{figure*}
Figure~\ref{fig1} shows a schematic of the experimental setup. The capacitor coil target consists of two parallel Cu foils (50~$\mu$m thick, $1.5 \times 1.5$~mm$^{2}$ in area) separated by 600~$\mu$m and connected by two parallel U-shaped Cu coils with a wire cross section of $100 \times 50$~$\mu$m$^{2}$. Each U-shaped coil is comprised of two straight wire segments, each 500~$\mu$m long, connected by a half-circular segment with a curvature radius of 300~$\mu$m. The separation between the two coils is 900~$\mu$m. Three short-pulse infrared laser beams from the LFEX laser, with a total energy of $\sim$750~J, a pulse duration of 1.5~ps, and a central wavelength of 1053~nm, were incident through a 400~$\mu$m-radius laser entrance hole in the front Cu foil and focused onto the back foil to a $\sim$100~$\mu$m-diameter focal spot at an incidence angle of $48^{\circ}$. The on-target laser energy was $\sim$430~J, corresponding to a peak laser intensity of $\sim4 \times 10^{18}$~W/cm$^{2}$.

Interferometry and proton radiography are used to diagnose the plasma density and magnetic field, respectively. A collimated 5-mm-diameter and 520-nm-wavelength probe laser is employed for interferometry. Since the probe beam is much larger than the plasma region, a self-referenced configuration is used in which one portion of the beam passes through the plasma while another portion propagates outside the plasma as a reference. After transmission through the target, these two portions are recombined to generate interference fringes.

Proton radiography is performed using an additional short-pulse infrared beam from the LFEX laser (0.3 kJ energy, 1.5 ps pulse duration, and 1053 nm central wavelength) incident on a 10-$\rm{\mu}$m-thick aluminum 1 mm $\times$ 1 mm foil to generate protons via the target normal sheath acceleration (TNSA) mechanism. The resulting proton beam traverses the central region between the two coil peaks and is recorded on stacks of radiochromic film (RCF). Each RCF layer corresponds to a specific proton energy, determined from calculations of proton energy deposition and the location of the Bragg peak within the film stack. This allows the interaction time to be inferred from the proton time of flight to the main target and the timing offset between the drive and proton-generation beams~\cite{Gao2012,Gao2013}. The proton beam axis is oriented at 45° with respect to the vertical direction. The proton source foil is positioned 3 mm away from the center of the coils, while the distance from the front surface of the RCF stack to the coil center is 30 mm.

\begin{figure*}[t]
\includegraphics[width=1\textwidth]{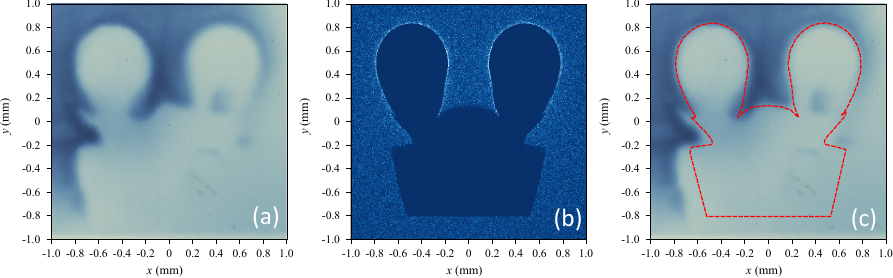}
\caption{\label{fig2}  Proton radiography of the capacitor--coil target. (a) Proton radiograph obtained with \(6.2 \pm 0.5\)~MeV protons at \(t = t_{0} + 0.59\)~ns. The color scale is not absolutely calibrated and qualitatively represents the proton flux, with darker regions indicating higher proton deposition. The coordinate axes are referenced to the target plane. 
(b) Synthetic proton radiograph calculated assuming a 29~kA current in each coil.
(c) Overlay of the synthetic radiograph on the experimental image. Dashed lines indicate the deflection contour derived from the synthetic proton radiograph shown in (b).
}
\end{figure*}

Figure~\ref{fig2} (a) shows a representative proton radiograph of the capacitor-coil target obtained at \(t = t_{0} + 0.59\)~ns, where \(t_{0}\) denotes the arrival time of the short-pulse laser irradiating the rear Cu plate. The proton energy was \(6.2 \pm 0.5\)~MeV. Strong proton deflection is observed in the vicinity of the coil, indicating the presence of intense electromagnetic fields. The deflection pattern is qualitatively consistent with magnetic-field--dominated deflection, as the proton displacement near the coil peaks is significantly larger than that near the straight coil legs~\cite{zhang2025}. Electric-field contributions to the proton deflection are expected to be small under these conditions, based on previous measurements and modeling of capacitor-coil targets.
\cite{Gao2025APL_CapacitorCoil,zhang2025}.

To quantify the electric current flowing through the coil, the experimental geometry was modeled using the charged-particle radiography module of PlasmaPy~\cite{plasmapy_community_2024_12788848} to generate synthetic proton radiographs for direct comparison with the experimental data. The magnetic-field distribution was calculated using the Biot--Savart law, assuming two U-shaped current paths with geometrical parameters taken from the fabricated target~\cite{zhang2025}. The proton source characteristics and radiography geometry were set to match the experimental conditions. Proton trajectories were traced through the calculated fields, with protons intersecting the target assumed to be blocked. A synthetic radiograph was then constructed by accumulating proton counts on the detector plane after propagation.

Figure~\ref{fig2} (b) shows the simulated proton radiograph corresponding to a current of 29~kA flowing through each coil. The simulated deflection pattern closely reproduces the key features observed in the experiment. An overlay of the simulation on the experimental image is shown in Fig.~\ref{fig2} (c), where the contour of the simulated deflection agrees well with the measured proton deflection boundary. Minor discrepancies near the plates are attributed to uncertainties in target fabrication and deviations between the idealized model geometry and the actual target. By varying the coil current in the simulations and comparing the resulting deflection contours with the experimental data, the current is estimated to be \(29 \pm 5\)~kA. The uncertainty in the estimated current is determined by identifying the range of currents for which clear deviations from the experimental deflection pattern become apparent. The measured current is lower than values reported in similar experiments\cite{Gao2025APL_CapacitorCoil} at the OMEGA-EP facility, which is attributed to differences in laser energy, pulse duration, and focal conditions between the two experiments.

\begin{figure*}[t]
    \centering
    \includegraphics[width=1\textwidth]{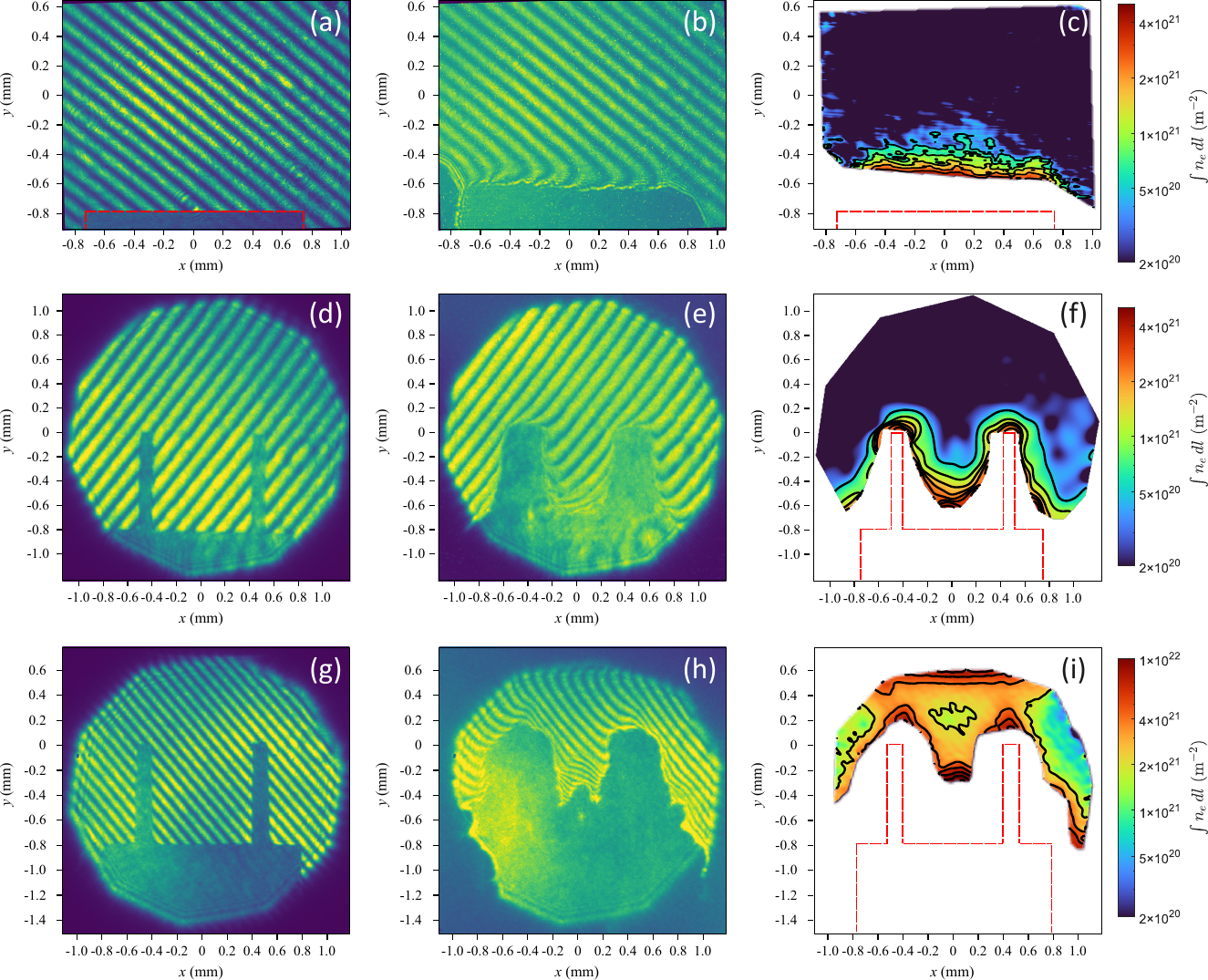}
    \caption{\label{fig3}
    Interferometry measurements for different target geometries at different times.
    (a-c) Measurements at $t = 1.0$~ns for a no-coil target consisting only of the back and front plates, with the coils removed.
    (d-f) Measurements at $t = 1.0$~ns and (g-i) at $t = 3.1$~ns after laser irradiation.
    Panels (a,d,g) show the reference interferograms acquired before the laser shot.
    Panels (b,e,h) show the interferograms during the shot.
    Panels (c,f,i) show the inferred line-integrated electron density maps. Density reconstruction is performed only in regions with clear interferometric fringes; areas without reliable fringe information are masked and excluded, resulting in a smaller reconstructed area than the corresponding interferograms. White regions indicate masked areas. Red dashed lines denote the original target position.
    }
\end{figure*}

Figure \ref{fig3} shows time-resolved interferometry measurements and the inferred line-integrated electron density for different target geometries. We used the Fourier-domain phase retrieval technique\cite{takeda1982fourier} by Interferometrical Data Evaluation Algorithms (IDEA) software\cite{HippReiterer2003_IDEA1.7} to extract the plasma-induced phase shift. The line-integrated density $\int n_e\, dl$ is calculated from the phase shift $\Delta\phi$ by
\begin{equation}
\int n_e\, dl = \frac{\lambda n_c}{\pi}\,\Delta\phi
\end{equation}
where $\lambda=$ 520 nm is the probe light wavelength and $n_c=4.14\times10^{27}\; \rm{m}^{-3}$ is the critical density of the probe beam. A phase retrieval error of approximately 0.2 rad can be present in the measured phase shift in some regions. The uncertainty in the inferred line-integrated electron density is therefore dominated by the phase retrieval error. A typical phase uncertainty of \(\sim 0.2\)~rad corresponds to an uncertainty of \(\sim 1.4 \times 10^{20}\,\mathrm{m^{-2}}\) in the line-integrated density, which also represents the minimum plasma density regime that can be reliably determined in our measurements. This level of uncertainty does not alter the qualitative spatial structure of the measured density distributions, as the inferred densities are substantially larger than the associated uncertainty.

Figures~\ref{fig3} (a-c) show measurements at $t = 1.0$~ns for a no-coil target consisting only of the back and front plates. The interferogram in Fig.~\ref{fig3} (b) exhibits fringe distortions near the front plate, and the reconstructed density map in Fig.~\ref{fig3} (c) shows measurable plasma density above the front plate in the region where the coil legs would normally be located. No significant density enhancement is observed near the coil-peak locations; however, the interferometric measurement is limited by a lower detection threshold, and plasma with densities below the threshold may already have expanded into this region.  We estimate the plasma expansion velocity using the position of the $\int n_e\,dl = 4\times10^{20}\,\mathrm{m^{-2}}$ contour. The laser interaction point is located at $y = -1.55$~mm, and at $t = 1.0$~ns this contour has propagated to $y \approx -0.5$~mm, corresponding to a propagation distance of $\sim 1.0$~mm. This implies an average expansion velocity $\sim 1000$~km/s, consistent with laser-ablation-driven expansion.

The coil-target case at the same time, $t = 1.0$~ns, is shown in Figs. 3 (d–f). The interferogram in Fig.~\ref{fig3} (e) exhibits pronounced fringe distortions localized near the coil peak positions. The reconstructed density map in Fig.~\ref{fig3} (f) reveals two high-density lobes spatially correlated with the coils, with peak line-integrated densities of $\sim (2$--$4)\times10^{21}\,\mathrm{m^{-2}}$. This demonstrates that the coil could also generate plasma and load near the coil and into the inside region.  The possible physical origins of this coil-localized plasma include plasma generation by strong inductive electric fields during the rapid current ramp-up, ohmic heating of the coil material, and radiative or energetic-particle heating driven by x-rays and energetic particles produced during laser ablation of the back plate.

At a later time $t = 3.1$~ns , shown in Figs.~\ref{fig3} (g-i), the density distribution extends over a larger volume and reaches higher values, with peak line-integrated densities up to $\sim 10^{22}\,\mathrm{m^{-2}}$, indicating substantial plasma accumulation in the inter-coil region. This behavior is consistent with continued plasma loading from both the coil-generated and ablation-driven sources.

These measurements provide direct evidence that plasma loading in laser-driven capacitor-coil targets arises from two distinct sources: plasma generated in the coil itself and plasma supplied by expansion from the laser interaction region. The presence of coil-generated plasma implies that the region surrounding the conductor is not vacuum, but instead filled with a conducting plasma that can modify the effective conductivity, current distribution, and magnetic-field topology near the coil, potentially affecting the magnitude and temporal evolution of the generated magnetic field. At the same time, ablation-driven plasma expansion provides an additional contribution to plasma loading in the inter-coil region, influencing the coupling between the externally generated magnetic field and secondary plasmas.

For application to laboratory studies of magnetic reconnection, these results highlight that the plasma environment surrounding capacitor-coil targets is dynamic. The target-generated plasma present at early times may precondition the reconnection region and the onset of current-sheet formation. Subsequent plasma loading can further modify the reconnection dynamics by changing density gradients and filling new plasma into the reconnection region. Incorporating time-resolved characterization of plasma density evolution is therefore important for interpreting the results of reconnection experiments.

When capacitor-coil targets are used to magnetize secondary plasmas, e.g., for fusion studies, the evolving plasma environment can directly influence experimental conditions. Plasma produced by the target can prefill the interaction region, alter background density and collisionality, and reduce magnetic-field penetration into secondary targets or plasmas. As a result, both the timing and magnitude of the applied magnetic field experienced by the secondary plasmas can differ from vacuum-field expectations, underscoring the importance of accounting for plasma evolution in target design, modeling, and experimental interpretation.

In summary, we have presented time-resolved interferometric measurements of the plasma density in laser-driven capacitor--coil targets irradiated by the University of Osaka LFEX laser. Proton radiography indicates a coil current of \(29 \pm 5\)~kA at \(t = 0.59\)~ns, confirming efficient current generation. The interferometry measurements reveal that plasma loading in the coil region arises from two distinct sources: plasma generated in the coil itself and plasma supplied by expansion from the laser interaction region. These measurements provide direct quantitative constraints on plasma loading and density distributions in capacitor--coil targets, which are essential for the interpretation and modeling of magnetized high-energy-density plasma experiments.

The authors thank the technical support staff at The University of Osaka (OU) for assistance with laser operation, target fabrication, and plasma diagnostics. 
The authors acknowledge the Research Fabrication Support Division of the University of Osaka Core Facility Center for assisting in the design/fabrication of several items. Yang Zhang thanks Yong Ma for helpful discussions in interferometry analysis.
This work was supported by the NASA Living with a Star Jack Eddy Postdoctoral Fellowship Program administered by UCAR's Cooperative Programs for the Advancement of Earth System Science (CPAESS) under award $\#$80NSSC22M0097US; by the U.S. Department of Energy the High-Energy-Density Laboratory Plasma Science program under Grant No. DE-SC0020103; by the Joint Usage/Research Center program of the Institute of Laser Engineering (ILE) at OU and Grants-in-Aid for Scientific Research (Nos. 25K17369, 23K03360, 22H00118, 22H01205, 22H01206, 22K03567, 21H04454, 20H00140, 20H01886, 17K05728, and 16H02245), "Power Laser DX Platform" as research equipment shared in the Ministry of Education, Culture, Sports, Science and Technology Project for promoting public utilization of advanced research infrastructure (Program for advanced research equipment platforms, grant number JPMXS0450300021); and by the Japan Society for the Promotion of Science Core-to-Core Program (grant number JPJSCCA20230003). 
The University of Osaka Honors Program for Graduate Schools in Science, Engineering, and Informatics supports Y. K., H. M. and R. Y.
The QLEAR fellowship program of OU partially supports J. D.

The data that support the findings of this study are available from the corresponding author upon reasonable request.

%---------------------------------------------------

%\bibliographystyle{1MAIN}
\bibliography{1MAIN}

\end{document}